\documentclass[traditabstract]{aa}
\usepackage{graphicx}
\usepackage{natbib}
\usepackage{amssymb}
\usepackage{amsfonts}
\usepackage{amsbsy}
\usepackage{amsmath}

\bibpunct{(}{)}{;}{a}{}{,}

\newcommand{\PFA}{P_{\rm FA}}
\newcommand{\PD}{P_{\rm D}}

\newcommand{\zerob}{\boldsymbol{0}}
\newcommand{\oneb}{\boldsymbol{1}}

\newcommand{\ab}{\boldsymbol{a}}
\newcommand{\bb}{\boldsymbol{b}}
\newcommand{\cb}{\boldsymbol{c}}
\newcommand{\eb}{\boldsymbol{e}}
\newcommand{\fb}{\boldsymbol{f}}
\newcommand{\gb}{\boldsymbol{g}}
\newcommand{\nb}{\boldsymbol{n}}

\newcommand{\ssb}{\boldsymbol{s}}

\newcommand{\xb}{\boldsymbol{x}}
\newcommand{\wb}{\boldsymbol{w}}
\newcommand{\zb}{\boldsymbol{z}}

\newcommand{\Cb}{\boldsymbol{C}}
\newcommand{\Hb}{\boldsymbol{H}}
\newcommand{\Lb}{\boldsymbol{L}}
\newcommand{\Mb}{\boldsymbol{M}}
\newcommand{\Pb}{\boldsymbol{P}}

\newcommand{\Xb}{\boldsymbol{X}}

\newcommand{\lambdab}{\boldsymbol{\lambda}}
\newcommand{\alphab}{\boldsymbol{\alpha}}
\newcommand{\deltab}{\boldsymbol{\delta}}
\newcommand{\sigmab}{\boldsymbol{\sigma}}
\newcommand{\Deltab}{\boldsymbol{\Delta}}

\newcommand{\Smatb}{\boldsymbol{{\mathcal S}}}
\newcommand{\Xmatb}{\boldsymbol{{\mathcal X}}}

\newcommand{\Cmatb}{\boldsymbol{{\mathcal C}}}

\newcommand{\Fmatb}{\boldsymbol{{\mathcal F}}}
\newcommand{\Gmatb}{\boldsymbol{{\mathcal G}}}
\newcommand{\Nmatb}{\boldsymbol{{\mathcal N}}}
\newcommand{\Pmatb}{\boldsymbol{{\mathcal P}}}

\newcommand{\Zmatb}{\boldsymbol{{\mathcal Z}}}

\begin{document}

   \title{Two modified ILC methods to detect point sources\\ in Cosmic Microwave Background maps}
   \subtitle{An approach with local background subtraction}

   \author{ Roberto Vio\inst{1}
   \and  Paola Andreani \inst{2,3} 
 	Elsa Patr\'icia R. G. Ramos\inst{2,4,5} \and Antonio da Silva\inst{4}}
   \institute{ Chip Computers Consulting s.r.l., Viale Don L.~Sturzo 82,
              S.Liberale di Marcon, 30020 Venice, Italy\\
              \email{robertovio@tin.it},
        \and
 	     ESO, Karl Schwarzschild strasse 2, 85748 Garching, Germany
         \and
               INAF-Osservatorio Astronomico di Trieste, via Tiepolo 11, 34143 Trieste, Italy\\              
		\email{pandrean@eso.org}
        \and
	     Centro de Astrof\'isica, Universidade do Porto, Rua das Estrelas, 4150-762 Porto, Portugal\\
             \email{eramos@astro.up.pt}      \\
 	     \email{asilva@astro.up.pt}       
        \and
	     Departamento de F\'isica e Astronomia, Faculdade de Ci\^encias, Universidade do Porto, Rua do Campo Alegre, 4169-007 Porto, Portugal
             }

\date{Received .............; accepted ................}

\abstract{
We propose two detection techniques that take advantage of a small sky area approximation and
are based on modifications of the {{\it internal linear combination}} (ILC) method, an approach widely used in Cosmology for the separation of the various components that contribute to the microwave background.
The main advantage of the proposed approach, especially in handling multi-frequency maps of the same region,  is that
it does not require the {{\it a priori}} knowledge of the spatial power-spectrum of either the CMB and/or the Galactic foreground. Hence, it is more robust, easier and more intuitive to use. 
The performance of the proposed algorithms is tested with numerical experiments that mimic the physical scenario expected for high Galactic latitude observations with the Atacama Large Millimeter/submillimeter Array (ALMA).}
\titlerunning{Point-Source Detection}
\authorrunning{R. Vio, \& P. Andreani}
\maketitle

\section{Introduction}

The detection of (extragalactic)  point-sources in experimental microwave maps is a critical step in the analysis of the Cosmic Microwave Background (CMB) maps.
Beside the specific interest related to
the construction of dedicated catalogues, if not properly removed, these sources can have adverse effects on the estimation of the power-spectrum and/or the test of Gaussianity of the CMB
component.
For this reason, this subject has been extensively considered in literature. Various detection techniques have been proposed but no 
general consensus about their real performances and properties has been reached. The reason is due to the different approaches adopted for dealing with the CMB and foregrounds (both extragalactic and Galactic) that make difficult the development of a general theory. Often such contributions are considered as a ``noise''  to be added to the instrumental noise. The various methods differ
in the statistical characteristics attributed to such noise as well as in the way to deal with it. 

Many efforts have been dedicated to the case of multiple frequencies maps of the same sky-area and
many algorithms have been proposed \citep[see][ and references therein]{her08, her12}. Apart from a recent Bayesian approach \citep{car09}, most of them belong to two broad classes of techniques. 
The first class makes use on the  {\it Neyman-Pearson} (NP)
criterion that consists of the maximisation of the {\it probability of detection} $\PD$ under the constraint that the {\it probability of false alarm} $\PFA$ (i.e., the probability of a false detection) does not exceed a fixed value
$\alpha$ \citep{kay98}. The resulting algorithms are an extension of the classic matched filter (MF). The second class is based on the constrained maximisation of the ``{\it signal-to-noise ratio}'' (${\rm SNR}$) of the source intensity with respect to the underlying background.
The constraints are chosen in such a way to improve detection. This class provides algorithms of ``{\it internal linear combination}'' (ILC) type. In Cosmology ILC is essentially  used 
for the separation of the various components contributing to the microwave sky emission \citep{eri04, hin07, vio08}. 
The main limitation of both classes is the requirement that the background is made of realisations of stationary stochastic processes with spatial spectra that, in addition, are supposed to be known. For example, on small patches of sky,
the Galactic contribution is modelled with stationary, in the case of NP algorithms Gaussian, processes with steep spectra (e.g. $1/f$ noise). These are rather rough assumptions. 
For this reason, in the context of future high spatial resolution observations
\citet{ram11} propose a method for high Galactic observations where the maps, almost free of the Galactic contamination, are linearly combined in such a way  that the resulting one is free of the CMB emission. 
Although such method has been successfully applied to the WMAP maps, it suffers from the drawback that it is unsuited for the detection
of point-sources with spectrum similar to that of the CMB or the Sunyaev-Zeldovich effects.
In this paper, we follow a different approach, based on two algorithms to be used in conjunction, that does not require the {\it a priori} knowledge of the spatial power spectra of the diffuse emissions due to either CMB or to the Galaxy
and at the same time it permits to deal with the above mentioned problem.
The basic consideration is that the detection of point sources is typically done on very small areas of sky
where both the CMB and the Galactic components can be very well approximated with low-degree two-dimensional polynomials.
 This is a more reasonable assumption than the ones mentioned above, i.e. emissions resulted from realisation of stochastic spatial processes.  
The performances of the algorithms is tested via numerical experiments based on simulated maps of high Galactic latitude that might be the area of interest of  
 CMB high spatial resolution observations.

\section{Formalization and  solution of the problem}

When looking for a point source in an area of sky, data can be modelled as two-dimensional discrete maps $\{ \Xmatb_i \}_{i=1}^{N_f}$, each of them containing
$N_p = N_{p_1} \times N_{p_2}$ pixels, corresponding to $N_f$ different observing frequencies (channels), with the form
\begin{equation} \label{eq:observed2}
\Xmatb_i = \Smatb_i + \Cmatb_i +  \Zmatb_i + \Gmatb_i + \Nmatb_i.
\end{equation} 
Here, $\Smatb_i$ is the contribution of the point-sources at the $i$th frequency, $\Cmatb_i$,  $\Zmatb_i$ and  $\Gmatb_i$ are the backgrounds due to CMB, extragalactic and Galactic emissions, respectively,
and  $\Nmatb_i$ is the instrumental noise. In this model, the contribution of the point-sources is assumed in the form
\begin{equation} \label{eq:point}
\Smatb_i = a_i \Fmatb,
\end{equation}
with ``$a_i$'' the intensity of the source at the $i$th channel. According to 
Eq.~(\ref{eq:point}), and without loss of generality, all the sources are assumed to have the same profile $\Fmatb$ independently of the 
observing frequency. In practical applications, this is not true. However, it is possible to meet this condition by convolving the images with an appropriate kernel with no remarkable consequences (see below). 

For computational reasons, that soon will become evident, it is useful to convert the two-dimensional model~(\ref{eq:observed2}) into the one-dimensional form
\begin{equation} \label{eq:observed1}
\xb_i = \ssb_i + \cb_i + \zb_i + \gb_i + \nb_i.
\end{equation}
Here, $\xb_i =  {\rm VEC}[\Xmatb_i]$, with  ${\rm VEC}[ \Hb]$ the operator that transforms a matrix $\Hb$ into a vector by stacking its columns one underneath the other. Something similar holds for the other quantities.

\subsection{Detection with ILC background removal} \label{sec:removal1l}
The main issue in detection problem when more maps of the same area are available at different observing frequencies, is how to handle them. One classical solution consists to linearly compose the maps
by means a set of weights $\wb=[w_1, w_2, \ldots, w_{N_f}]^T$ in such a way that it is possible to work with a single map given by
\begin{equation}
\xb = \Xb \wb,
\end{equation}
where $\Xb = [ \xb_1, \xb_2, \ldots, \xb_{N_f}]$ is a $N_p \times N_f$ a matrix.
The question is how to fix such weights. In accomplishing such a task, it is necessary to take into account that there is some {\it a priori} information about the various components in Eq.~(\ref{eq:observed2}). 
In particular:
\begin{itemize} 
\item For each observing frequency $i$, the spectra of $\Cmatb_i$ and $\Zmatb_i$ are known with good accuracy. This is the case for instance of the CMB and the Sunyaev-Zeldovich (SZ) effects, both thermal and kinetic;\\
\item $\Cmatb_i$ and $\Gmatb_i$ have a diffuse spatial distribution with a spatial scale much greater than that of the point sources; \\
\item Noises $\Nmatb_i$ can be reasonably assumed as given by the realisation of independent Gaussian white-noise processes with standard deviation $\sigma_{\Nmatb_i}$.
\end{itemize}

We are interested in exploring the situation in which in addition to the CMB the extragalactic component consists of secondary anisotropies of the CMB.
Here we consider only the SZ effects which are the strongest ones in galaxy clusters, groups of galaxies and in protoclusters \citep{bir99} and whose spectral shape is well known.
The first point implies that the contribution $\bb$ to $\xb$ of the CMB and of the SZ components  can be obtained from
\begin{equation}
\bb = \Mb [\cb, \zb],
\end{equation}
where $\cb$ and $\zb$ are templates for CMB and SZ components (i.e. maps that do not depend on frequency) and $\Mb$ is a $N_p \times N_e$ {\it mixing matrix}. In the present context, $N_e = 2$ since the kinetic SZ emission
has the same spectrum of CMB. For this reason, from now on, with $\cb_i$ we will indicate the CMB plus the kinetic SZ emission.  The second point implies that within a small area centred at a point source the CMB and the Galactic emissions
vary very little. This suggests that, for any sub-map $\Xmatb_i(j,k)$ with $-N_j \le j \le N_j$ and $-N_k \le k \le N_k$  ($N_j \ll N_{p_1}$ and  $N_k \ll N_{p_1}$),
these emissions can be safely approximated by a low-degree, two-dimensional polynomial of degree $m$
\begin{equation}
\Pmatb_m (j,k) = \sum_{l=0}^m \alpha_l (j^q k^r); \qquad q + r \le l,
\end{equation}
where $\{ \alpha_l \}$ are real coefficients whereas $q$ and $r$ are integer numbers permuted accordingly.

If for detection one adopts the criterion of the {\rm SNR} maximisation, all these considerations suggest a model where the weights, the point source intensity and the parameters of the approximating two-dimensional polynomial
in $\xb$ are optimised simultaneously, i.e.
\footnote{We recall that the functions ``$ \underset{x}\arg\min F(x)$'' and 
``$ \underset{ x }\arg\max F(x)$''
provide the values of $x$ for which the function $F(x)$ has the smallest and greatest value, respectively.},
\begin{equation} \label{eq:problem1}
R(\wb, \cb,\lambdab) =  \underset{ \wb, \cb, \lambdab}{\arg\min}  \left[ \Vert ( \Xb \wb - \Lb \cb) \Vert^2 + \lambdab^T ([\ab, \Mb]^T \wb - \eb_1) \right] .
\end{equation}
Here, ``$\Vert . \Vert$'' indicates Euclidean norm, $\ab = [a_1, a_2, \ldots, a_{N_f}]^T$ is an array containing the emission spectrum of the source to detect, $\cb=[a, \alphab^T]^T$ is an array with size $N_c=[(m+1) (m+2)/2] +1$ with ``$a$'' 
the amplitude of the source in the linearly composed 
sub-map of $\Xmatb$ and $\alphab$ the coefficients of the two-dimensional polynomial.  $\Lb$ is a $N_p \times N_c$ matrix with the form $\Lb = [\fb, \Pb]$ where $\fb = {\rm VEC}[\Fmatb]$ and $\Pb$ the $N_p \times (N_c-1)$ matrix
that results from the least-square fit of a two-dimensional polynomial. For example, when  $m=1$, $\Pb = [\deltab_1, \deltab_2, \oneb]$ whereas, when  $m=2$, 
$\Pb = [\deltab_1 \odot \deltab_1, \deltab_2 \odot \deltab_2, \deltab_1 \odot \deltab_2, \deltab_1, \deltab_2, \oneb]$, where ``$\odot$'' represents the {\it element-wise}
matrix multiplication (Hadamard product), $\oneb$ is a vector of ones  and $\deltab_1 = {\rm VEC}[\Deltab_1]$, $\deltab_2 = {\rm VEC}[\Deltab_2]$ where $\Deltab_1$ is a matrix with $2 N_j+1$ identical columns 
$[-N_k, -N_k+1, \ldots, 0, \ldots, N_k-1, N_k]^T$ whereas $\Deltab_2$ is a matrix with $2 N_k+1$ identical rows $[-N_j, -N_j+1, \ldots, 0, \ldots, N_j-1, N_j]$.
Finally $\lambdab$ is a $N_e+1$ array of {\it Lagrange multipliers} whereas $\eb_1$ is a $N_e + 1$  array of zeros except for the first element that is ``$1$''. With this model the quantity
${\rm SNR} = (\ab^T \wb)^2 /  \Vert ( \Xb \wb - \Lb \cb) \Vert^2$ is maximized under the constraint that $\ab^T \wb=1$ (i.e., preservation of the source amplitude) and $\Mb^T \wb = [0, 0]$ (i.e., the CMB and SZ components are zeroed).  

The basic idea behind this method, that we call {\it modified multiple} ILC (MMILC), is that, if in the centre of the selected sub-map there is a point-source, then the value of ``$a$'' should exceed a threshold due to  noise. 
After some algebra, one obtains that the solution of problem~(\ref{eq:problem1}) is given by the system of equations
\begin{equation} \label{eq:solution1}
\left( \begin{array}{ccc}
+2 \Cb_{XX} & -2 \Cb_{XL} & \Mb^T \\
 -2 \Cb_{XL}  & +2 \Cb_{LL} & \zerob \\
\Mb & \zerob & \zerob
\end{array} \right) 
\left( \begin{array}{c}
\wb \\
\cb \\
\lambdab
\end{array} \right) =
\left( \begin{array}{c}
\zerob \\
\zerob \\
\eb_1
\end{array} \right),
\end{equation}
where $\Cb_{XX} =  \Xb^T \Xb$, $\Cb_{XL} = \Xb^T \Lb$ and $\Cb_{LL}=\Lb^T \Lb$.
The explicit solution for $\wb$, $\cb$ and $\lambdab$ is not difficult to obtain but it produces rather complicated expressions. Hence, the numerical solution is more advantageous.
One interesting characteristic of solution~(\ref{eq:solution1}) is that it does not require the knowledge of the noise level of each map, a quantity that often can be only roughly estimated.

When searching for point sources in a given set of maps, the procedure consists in fixing the size $(2 N_j +1) \times (2 N_k +1)$  of a window that is made to slide, pixel by pixel, across the area of interest. At the end of this
procedure a single map is obtained containing the estimated values of ``$a$'' for each pixel. Now,  the question is to fix the detection threshold below which a given value of ``$a$'' is supposed
to be due only to noise.
In this respect, the direct use of solution~(\ref{eq:solution1}) is difficult. For this reason, two different procedures are suggested:
\begin{enumerate}
\item $a = 0$ if $a \le k \sigma_{{\rm L}}$, where $k$ is a constant factor (typically $k=4,5$), $\sigma_{{\rm L}} = \Vert \sigmab^T_{n}  \wb \Vert \sqrt{(\Lb^T \Lb)^{-1}_{1, 1}}$,  $ \sigmab_{n} =  
[ \sigma_{n_1},  \sigma_{n_2}, \ldots,  \sigma_{n_{N_f}}]^T$ and
$(\Lb^T \Lb)^{-1}_{1, 1}$ is the first entry of matrix  $(\Lb^T \Lb)^{-1}$. This operation corresponds to estimate the standard deviation $\sigma_a$ of ``$a$'' for a fixed
$\wb$. Such an approach has the advantage that the matrix $(\Lb^T \Lb)^{-1}$  can be computed only once since it is the same for all the sub-maps. But, it has the disadvantage that the standard deviations of the noises 
$\{ n_i \}$ are to be known in advance; \\ 
\item  $a=0$ if $a \le k \sigma_{{\rm map}}$, where again $k$ is a constant factor and $\sigma_{{\rm map}}$ is the standard deviation of the entries in the final map. This is an unsophisticated approach, however it has the advantage
that does not require the standard deviation of the noise in each sub-map, a quantity usually only roughly known.
\end{enumerate}
Perhaps an advisable procedure consists in using both methods and to check for differences.

\subsection{Detection without ILC background removal}

The MMILC detection procedure is potentially quite effective, however it suffers of two main drawbacks:
\begin{enumerate}
\item In order to remove the CMB and SZ components, one or more of the weights in $\wb$ have to be negative. As a consequence, since in the final map $a = \ab^T \wb$ and $\sigma_{{\rm map}} = 
\Vert \sigmab^T_{n}  \wb \Vert$,  ``$a$'' is given by the sum of positive as well negative values whereas $\sigma_{{\rm map}}$ is given by the sum of positive values only. In other words, the background subtraction reduces
 the ${\rm SNR}$ with respect to a simple sum of the maps. The situation worsens when the emission of a point source has a spectrum similar to that of CMB or of SZ since ``$a$'' will tend to zero; \\
\item If $\ab$ is an array such that $\Mb \ab = \zerob$, i.e. $\ab$ belongs to the {\it nullspace} of $\Mb$  (i.e., it is given by the linear combination of the column of $\Mb$) then the
system~(\ref{eq:solution1}) does not have any solution because the constraints $\ab^T \wb =1$  and $\Mb^T \wb = \zerob$ become incompatible.
\end{enumerate} 
For this reason, in order to detect point sources with $\ab$ belonging to the {\it nullspace} of $\Mb$, the above procedure has to be adapted to work without the ILC removal of the CMB and SZ components. More specifically,
problem~(\ref{eq:problem1}) is converted into
\begin{equation} \label{eq:problem2}
R(\wb, \cb, \lambda) =  \underset{ \wb, \cb, \lambda}{\arg\min}  \left[ \Vert ( \Xb \wb - \Lb \cb) \Vert^2 + \lambda (\ab^T \wb - 1) \right] ,
\end{equation}
with solution given by
\begin{equation} \label{eq:solution2}
\left( \begin{array}{ccc}
+2 \Cb_{XX} & -2 \Cb_{XL} & \ab \\
-2 \Cb_{XL} & +2 \Cb_{LL} & \zerob \\
\ab^T & \zerob^T & 0
\end{array} \right) 
\left( \begin{array}{c}
\wb \\
\cb \\
\lambda
\end{array} \right) =
\left( \begin{array}{c}
\zerob \\
\zerob \\
1
\end{array} \right).
\end{equation}
With this method, that we call {\it modified} ILC (MILC), the CMB  is not removed through the use of the mixing matrix $\Mb$. However the fact that this is a component with diffuse spatial distribution makes us hope that it can be removed 
through the polynomial
approximation of the background. As a consequence, in the final map the only contribution beyond that of the point sources is  the SZ (both thermal and kinetic). Unfortunately this is an unavoidable problem. Without further
information it is impossible to separate an SZ emission with point like spatial distribution from a genuine point source.

\section{Practical uses}
 
\begin{figure*}
        \resizebox{\hsize}{!}{\includegraphics{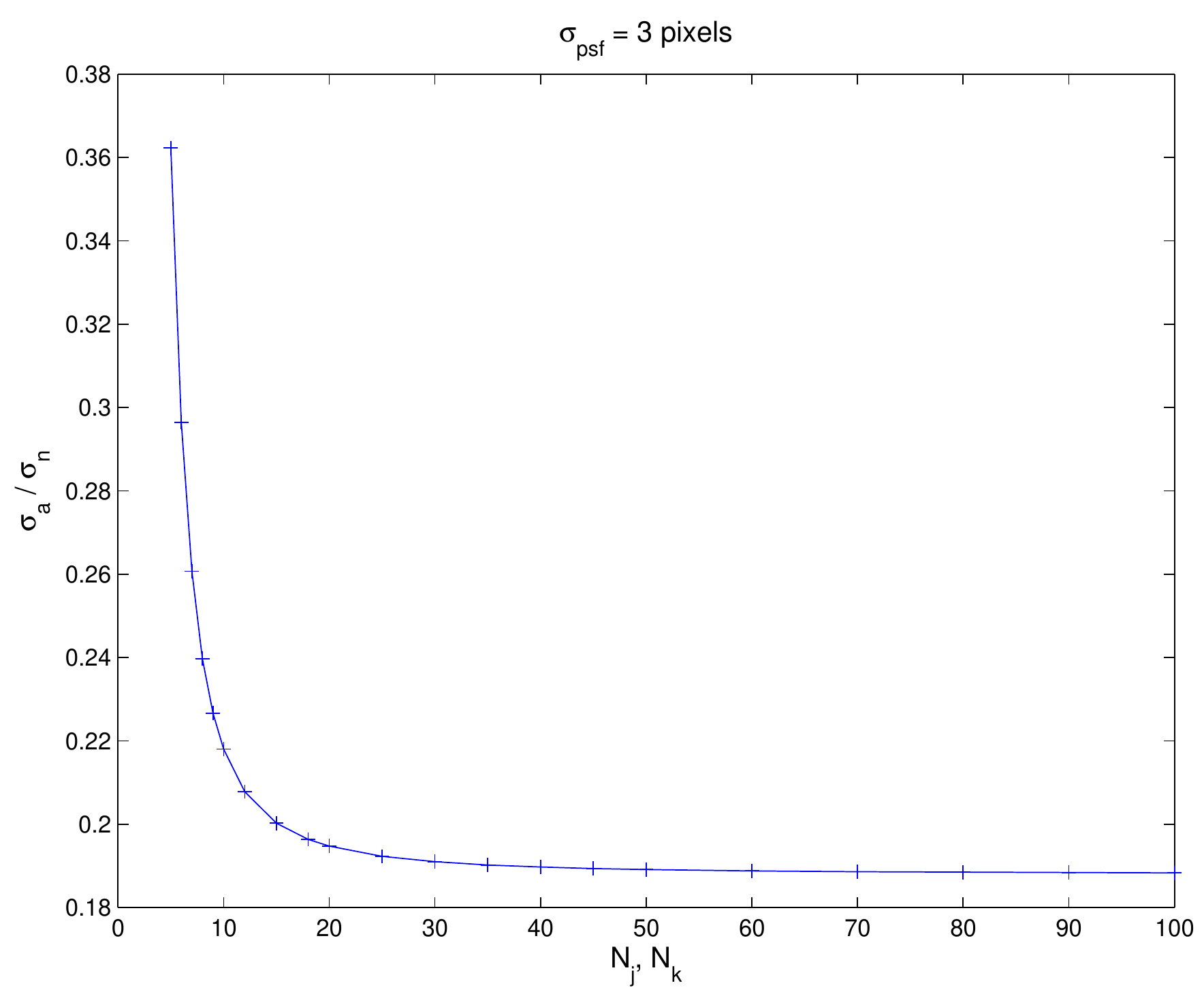}}
        \caption{Standard deviation $\sigma_a$ of the estimated intensity $a$  as provided by MILC in the case of a point source, with Gaussian profile with dispersion $\sigma_{\rm psf}$ equal to $3$ pixels, as a function of the sizes $N_j = N_k$ of the searching sub-map. Here, a single map is considered with a background given by a  two-dimensional one degree polynomial, instrumental noise is Gaussian and white with standard deviation $\sigma_n$, The true value of ``$a$'' 
is $1$ in unit of $\sigma_n$.}
        \label{fig:area1}
\end{figure*}

\begin{figure*}
        \resizebox{\hsize}{!}{\includegraphics{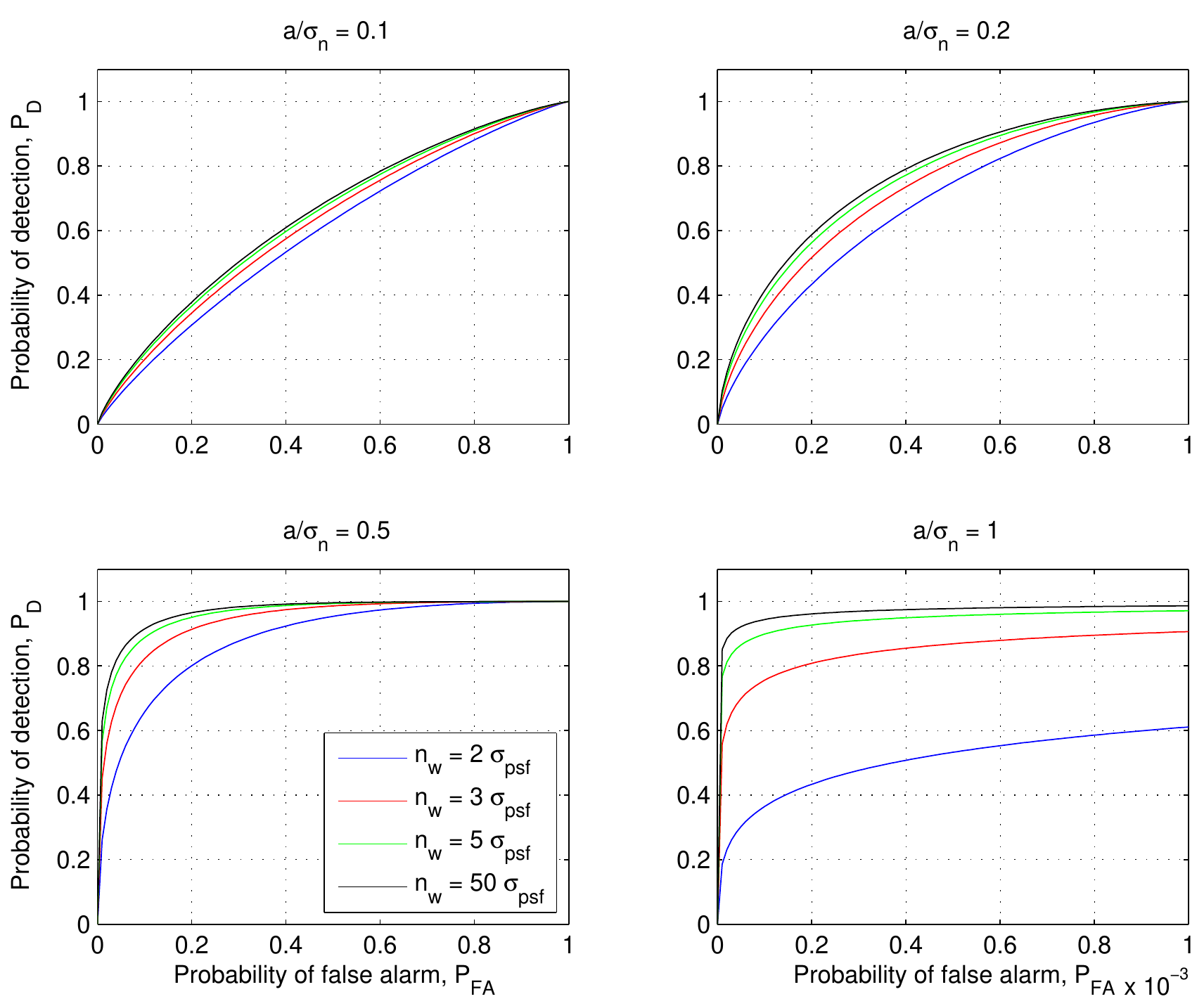}}
        \caption{Relationship between the {\it probability of detection}, $\PD$, vs. the {\it probability of false alarm}, $\PFA$ for the case shown in Fig.~\ref{fig:area1} but for different values of the ratio $a / \sigma_{n}$. Note the different
         scale used for the abscissa in the bottom-right panel.}
        \label{fig:area2}
\end{figure*}

In this section we discuss some practical problems and how they can be addressed. The first is related to the degree $m$ of the polynomial used to approximate the background. We need to take into account that, 
when two polynomials of degree $m_1$ and $m_2$ with
$m_1 > m_2$ are summed together, a polynomial with degree $m=m_1$ is obtained. Hence, $m$ is fixed by the diffuse component that requires the largest degree of the approximating polynomial.
If the sky patches are small, it can be reasonably expected that a first degree polynomial is a good choice.
The second question is related to the sizes $N_j$ and $N_k$ of the sub-map where to
test for the presence of a point source. Two competing requirements raise: on the one hand $N_j$ and $N_k$ must be as large as possible to reduce errors in the estimation of the polynomial parameters, on the other hand, a small size implies that the polynomial approximation for the background is preciser and a lower probability is expected with respect to the area where some other sources are present. For illustrative purposes, Fig.~\ref{fig:area1} shows 
the standard deviation $\sigma_a$ of the estimated intensity $a$ as provided by MILC in the case of a point source with a Gaussian profile and a dispersion $\sigma_{\rm psf}$ equal to $3$ pixels. A single map is considered where the background is given by a 
two-dimensional one degree polynomial, a Gaussian and white instrumental noise with standard deviation $\sigma_n$, and $N_{p_1} $, $N_{p_2}$ are increased. The true value of $a$ is $1$ in unit of $\sigma_n$.
The decrease of $\sigma_a$ is evident.  Figure~\ref{fig:area2} shows the relationship between $\PD$ and $\PFA$ for different values of the ratio $a / \sigma_n$. From these figures it is clearly shown that $N_j$, $N_k$ lying in the range $3 \sigma_{\rm psf}$-$5 \sigma_{\rm psf}$ is a reasonable compromise.

Another issue is related to the fact that in practical applications the width of the PFSs changes with observing frequency. Widespread practice is to convolve maps with a suited kernel function in order to get a common spatial resolution at all frequencies. This operation has the beneficial effect to reduce the standard deviation of the instrumental noise but at the same time it introduces a spurious spatial correlation in it. Actually, even if neglected, this latter
is not critical since  MILC and MMILC are linear techniques and the only consequence is a slight reduction of their efficiency. In other words, given the above mentioned reduction of the standard deviation of the noise
this spurious correlation is of secondary importance. This is especially true if one takes into account that there are other and more important approximations that make the analysis of data less rigorous (e.g., often the level of instrumental 
noise is only roughly known).

As a final comment, both MMILC and MILC work optimally only for a specific emission spectrum $\ab$, a limitation common to many other detection techniques. 
But this is not critical. It is sufficient to apply the detection algorithm each time with a different value of $\ab$.
This is made possible by the fact that MMILC and MILC are quite fast algorithms since they require the numerical solution of systems containing no more than a couple of tens of linear equations.  

\section{Numerical experiments}

In order to support the arguments presented above, here we present some numerical experiments with simulated maps at high Galactic latitude (where the Galactic contamination is negligible) that is the region of interest for future CMB experiments.

\begin{figure*}
        \resizebox{\hsize}{!}{\includegraphics{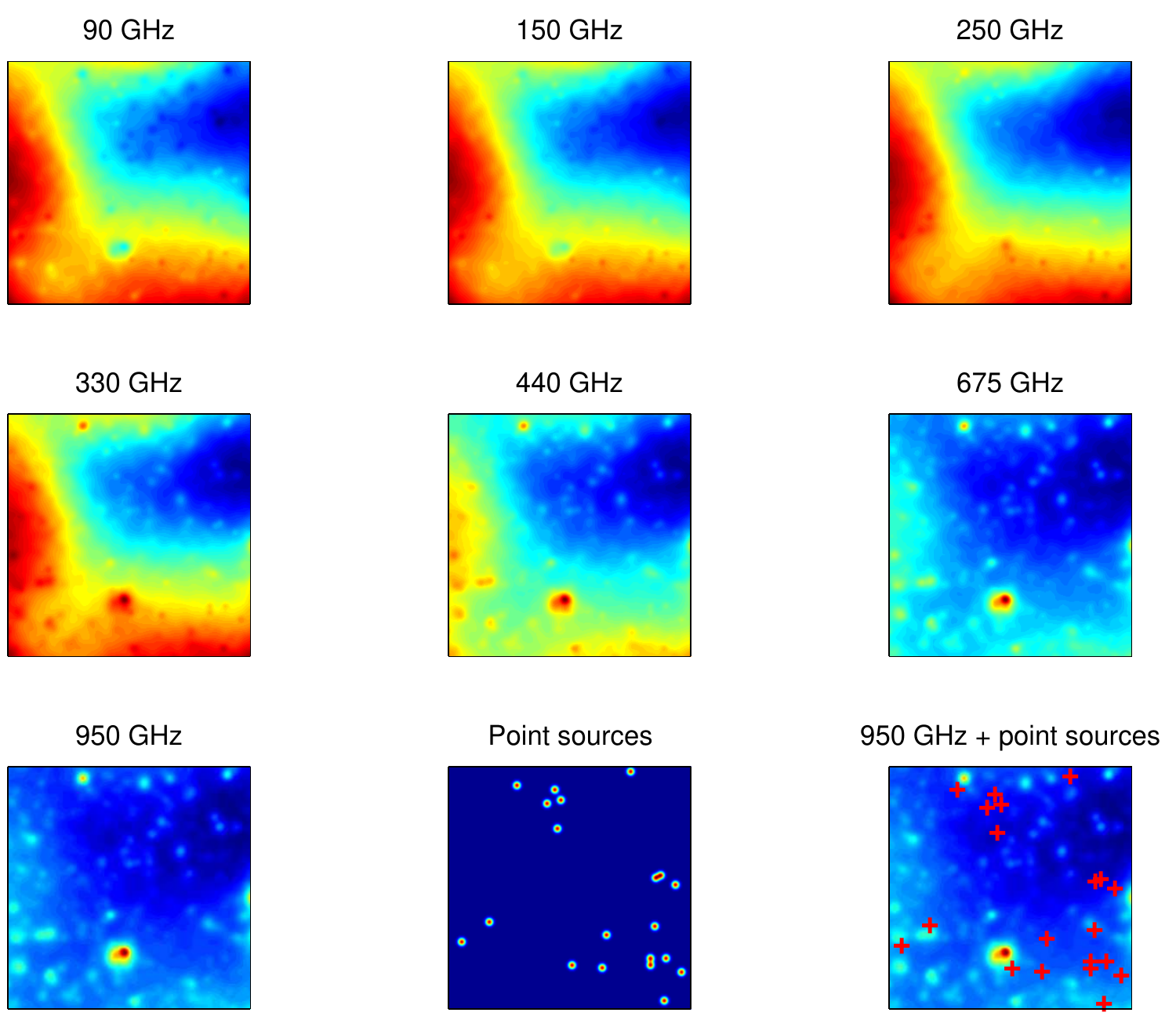}}
        \caption{Noise free maps simulating a high Galactic declination area of sky at the ALMA observing frequencies. $20$ randomly distributed point sources with the same intensity have been added. The point source intensity has been set to $1.7$ times the
        standard deviation of the noise (see next figure). In this way their spectrum mimics that of CMB + SZ (see text). The PSFs are assumed to be Gaussian with a standard deviation of $3$ pixels.  The two bottom-right panels show the simulated point sources
        and their position on the $950~{\rm GHz}$ map, respectively.}
        \label{fig:detection1a}
\end{figure*}

\begin{figure*}
        \resizebox{\hsize}{!}{\includegraphics{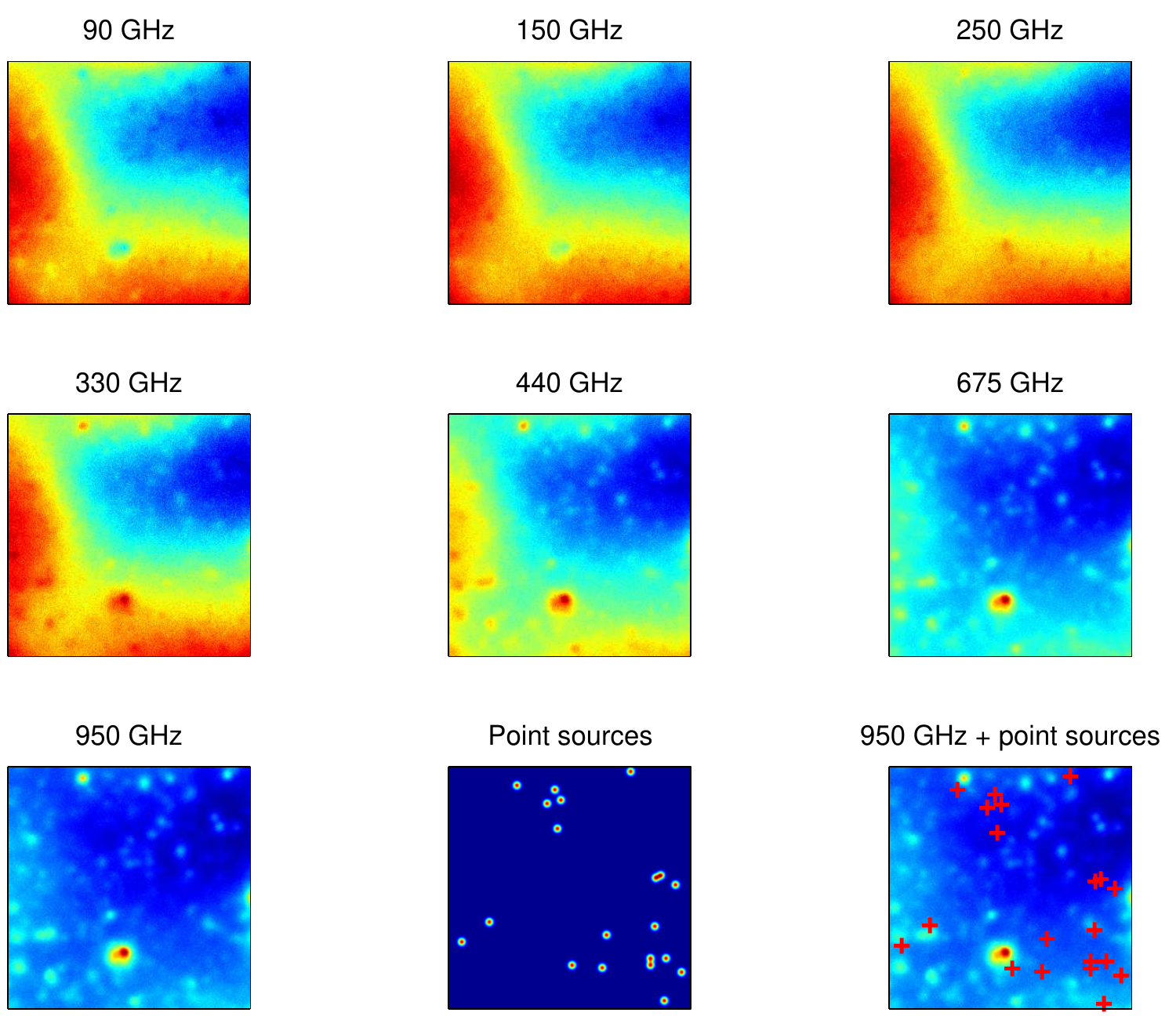}}
        \caption{Noisy version of the maps in Fig.~\ref{fig:detection1a}. Noise is Gaussian-white with standard deviation set to $0.12$ time the standard deviation of the values in the corresponding  noise free maps. All of the point sources have the same
         intensity set to $1.7$ times the standard deviation of the noise. The two bottom-right panels show the simulated point sources and their position on the $950~{\rm GHz}$ map, respectively. }
        \label{fig:detection1b}
\end{figure*}

\begin{figure*}
        \resizebox{\hsize}{!}{\includegraphics{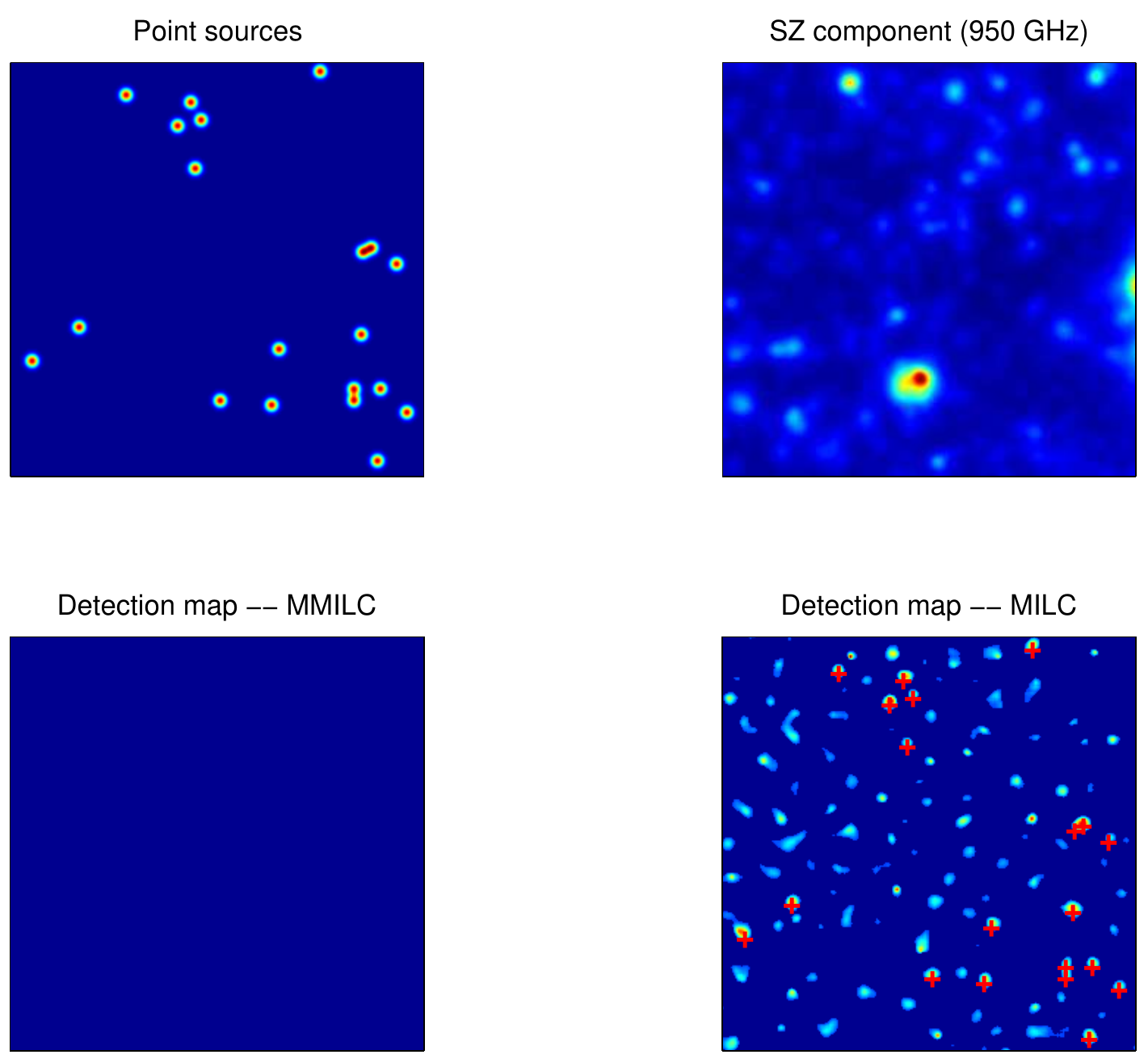}}
        \caption{Results provided by MILC and MMILC when applied to the maps in Fig.~\ref{fig:detection1b}. The detection threshold has been set to  $5 \sigma_{\rm L}$ (see text) and the background has been approximated with a
         two-dimensional polynomial of degree one. The bottom left panel shows clearly that MMILC is not able to retrieve point sources in this case. This happens because the point surges spectra has a frequency-dependence similar to that of the CMB and SZ and therefore the subtraction process gets rid of all of them. The bottom right panel shows that MILC, on the contrary, retrieve all sources and the SZ point-like emissions, because it subtracts the underlying diffuse component with the polynomial approximation.}
        \label{fig:detection1c}
\end{figure*}

\begin{figure*}
        \resizebox{\hsize}{!}{\includegraphics{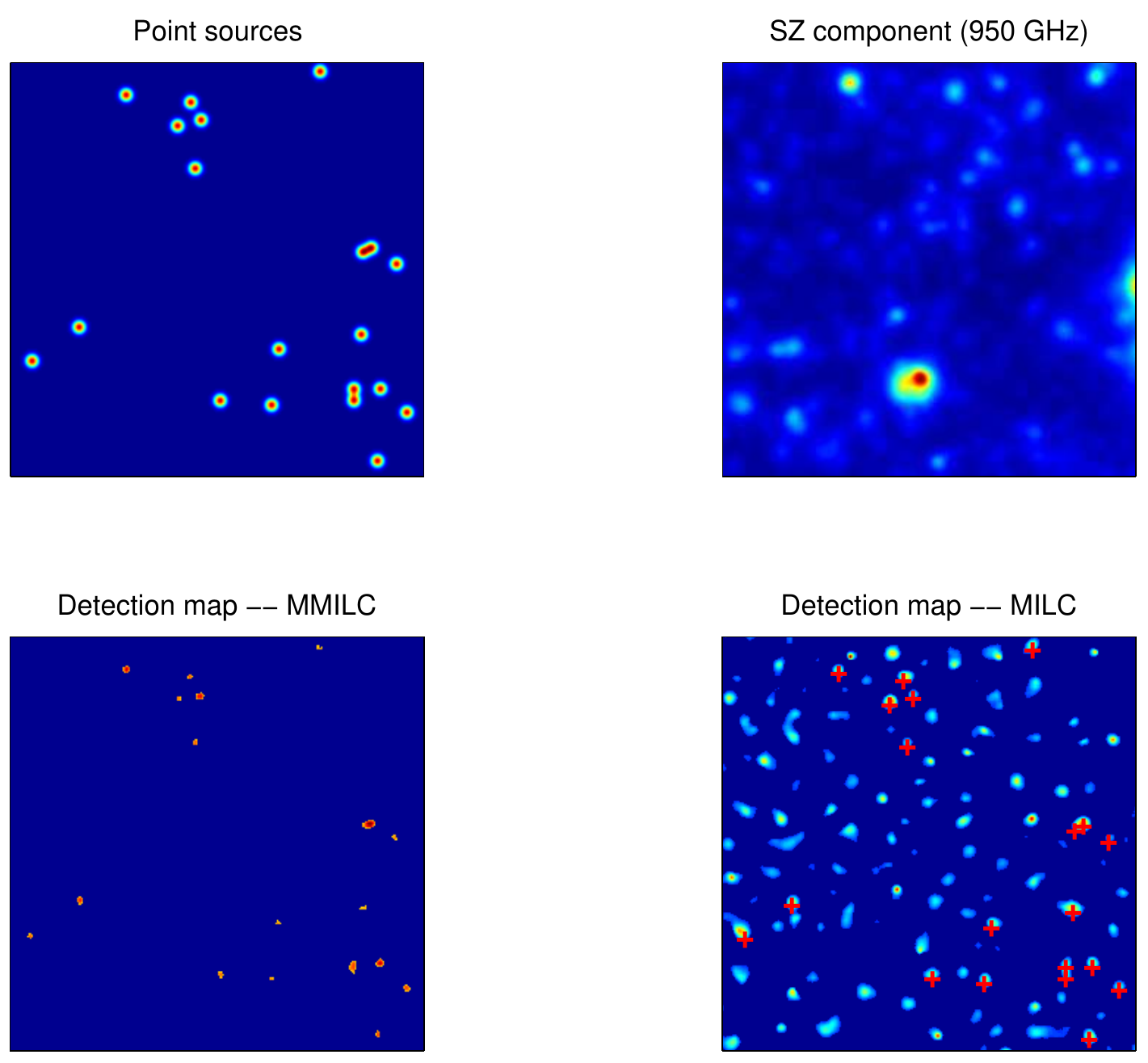}}
        \caption{As in Fig.~\ref{fig:detection1c} with the only difference that the intensity of the sources at $90~{\rm GHz}$ is set to zero. With this constrain both MMILC and MILC are able to
        detect all the point sources and get rid of the SZ point-like emissions.}
        \label{fig:detection2c}
\end{figure*}

We produced small sky patches of $0.86 ~{\rm deg}^{2}$ at $3''$ angular resolution with several components, 
namely, the CMB and the  Sunyaev-Zel'dovich effects (SZ), both kinetic and thermal. 
To produce these maps we used Hydrodynamic/N-body simulations with cosmological parameters consistent 
with WMAP parameters for a flat Universe and standard $\Lambda$ CDM model, with an equation of state 
for the dark energy component of $w=-1$. The adopted present time density parameters expressed in terms 
of the critical density are $(\Omega_{\rm cdm},\Omega_{\Lambda},\Omega_b)=(0.256, 0.7, 0.044)$, a 
dimensionless Hubble constant of $h=0.71$ and a mean CMB temperature of $T$=2.725 K. It is assumed 
adiabatic initial conditions, a spectral index of $n_s=1$ and full reionisation at redshift $7$. 
For the present epoch we considered a normalisation power spectrum of $\sigma_{8}=0.9$ 
and a shape parameter of $\Gamma=0.17$.  
The CMB component is produced with the CAMB code \citep{lew00} 
to obtain the linear CMB power spectrum. The full-sky CMB temperature anisotropy map was generated 
with the HEALPix software \citep{gor05} with ${\rm Nside}=8192$. From this map it was extracted a small sky 
region with an area of about 0.86 deg$^{2}$ around the equator, projected in a squared map.
Details about the simulations of the SZ effect components can be found in \citep{das01, ram12}. 
The frequencies chosen were $90$, $150$, $250$, $330$, 
$440$, $675$ and $950~{\rm GHz}$ which correspond to the receiver bands of ALMA.
All components were co-added resulting in a final map, $\Delta I_{\rm CMB+SZ}/I$, with pixel size of 3 arcsec.
We use the central part of the maps ($300 \times 300$ pixels) and convolve them for each frequency with a Gaussian PSF with dispersion of $3$ pixels. To each map a white-noise process with standard deviations $\sigma_{n_i}$ set to $0.12$ time the standard deviation of the values of 
map itself has been also added. Finally $20$ point sources randomly distributed have been included with $a_i = 1.7 \sigma_{n_i}$.  In this way, maps with the same {\rm SNR} are obtained.
The values of $\sigma_{n_i}$ and $a_i$  have been arbitrarily chosen in such a way to test algorithms under very bad operational conditions but at the same time to obtain
stable results (i.e. with different realizations of the noise process almost all the sources are correctly detected with no false detections). The simulated experimental scenario corresponds to
an adverse situation of rather low {\rm SNR} and, since $\sigma_{n_i}$ increases with frequency, with a spectrum $\ab$ that mimics that of the CMB plus SZ background (i.e. $\ab$ is close to the {\it nullspace} of $\Mb$, or $\Mb \ab \approx \zerob$).
Figures~\ref{fig:detection1a}-\ref{fig:detection1b} show the noise free and the noisy version of such maps, respectively. From these figures it is evident that most of the point sources are not even visible and that are by far exceeded by the SZ point-like emission.  Figure~\ref{fig:detection1c} shows the results obtained with MILC and MMILC using a detection
threshold $5 \sigma_{\rm L}$ and a background approximated by a two-dimensional first degree polynomial. As expected, the MMILC does not work. On the other side, MILC
has effectively removed the CMB components and correctly detected all the point-sources in the map. However, many of the SZ point-like emission is also present. The situation greatly improves when 
the value of amplitude of the point sources at $90~{\rm GHz}$ is set to zero. In this way $\Mb \ab \ne \zerob$, i.e. the degeneracy of the similarity of CMB and SZ kinetic spectra is broken.
Similar results are obtained setting the flux to zero at another frequency.
 As visible in Fig.~\ref{fig:detection2c}, while MILC provides almost the same results as before, MMILC is able to correctly detect all the point sources as well to completely remove the CMB and SZ contamination. A point to stress is that, in the present high Galactic latitude experiment  the diffuse Galactic component is set to zero. Hence, a simple ILC algorithm could be used (i.e.
without the subtraction of the polynomial approximation of the background). However, the use of MMILC permits to test this algorithm for higher level of noise.
 
\section{Summary and conclusions}

In this paper two different modifications of the {\it internal linear combination} algorithm, the {\it modified} ILC (MILC) and the {\it modified multiple} ILC (MMILC), have been presented to detect
point sources in multifrequency CMB maps. The remarkable property of the proposed algorithms is that they do not require the {\it a priori} 
knowledge of the statistical characteristics of the diffuse sky background since this is assumed to be locally approximate to a low-degree two dimensional polynomial. In fact,
the methods available in literature have the drawback to assume this component as the realisation of stationary, often Gaussian, two-dimensional stochastic processes, with known spatial power-spectrum, which is not a realistic approximation. The two proposed algorithms used in conjunction are effective in the detection of point sources independently of the spectral characteristics of their emission.
Their potential performance has been tested via numerical experiments.

\begin{acknowledgements} 
E. P. Ramos is supported by grant POPH-QREN-SFRH/BD/45613/2008, from FCT (Portugal). 
E. P. Ramos and R. Vio thank ESO for its hospitality and support through the DGDF funding programme.
\end{acknowledgements}

\clearpage


\begin{thebibliography}{}
\bibitem[Birkinshaw (1999)]{bir99} Birkinshaw M., 1999, Phys. Rep., 310, 97
\bibitem[Carvalho et al. (2009)]{car09} Carvalho, P., Rocha, G. and Hobson, M.P. 2009, MNRAS, 393, 681
\bibitem[da Silva et al. (2001)]{das01} da Silva A. J. C., Barbosa D., Liddle A. R., \& Thomas, P. A. 2001, MNRAS, 326, 155
\bibitem[Herranz and Sanz (2008)]{her08} Herranz, D. and Sanz, J.L. 2008, IEEE Journal of Selected Topics in Signal Processing, 2, 727
\bibitem[Herranz et al. (2012)]{her12} Herranz, D., Arg\"ueso, F. and Carvalho, P. 2012, Advances in Astronomy, {\it in print} 
\bibitem[Eriksen et al. (2004)]{eri04} Eriksen, H.K., Banday, A.J., G\'orski, K.M., \& Lilje, P.B. 2004, \apj, 612, 633
\bibitem[G\'orski et al. (2005)]{gor05} G\'orski, K. M., Hivon, E., Banday, A. J., Wandelt, B. D., Hansen, F. K., 
Reinecke, M., \& Bartelmann, M. 2005, \apj, 622, 759
\bibitem[Hinshaw et al. (2007)]{hin07} Hinshaw, G., et al. 2007, \apjs, 170, 288
\bibitem[Kay (1998)]{kay98} Kay, S. M. 1998, Fundamentals of Statistical Signal Processing: Detection Theory (London: Prentice Hall)
\bibitem[Lewis et al. (2000)]{lew00} Lewis, A., Challinor, A., \& Lasenby, A. 2000, \apj, 538, 473
\bibitem[Ramos et al. (2011)]{ram11} Ramos, E.P.R.G., Vio, R. and Andreani, P. 2011, A\&A, 528, A75
\bibitem[Ramos et al. (2012)]{ram12} Ramos, E. P. R. G., da Silva, A., J. C., Liu, G.C., 2012, ApJ, {\it in preparation}
\bibitem[Vio and Andreani. (2008)]{vio08} Vio, R. and  Andreani, P. 2008, A\&A, 487, 775
\end{thebibliography}
\end{document}